\documentstyle[12pt,epsf]{article}

\begin{document}
\hfill THU-93/21
\vspace*{5cm}
\begin{center}
{ 
\Large \bf RELATIVISTIC TWO-BODY BOUND-STATE  \vspace{0.1cm}
CALCULATIONS 
BEYOND THE LADDER  \\ \vspace{0.1cm}
APPROXIMATION.}\\
\vspace{1.0cm}
Taco Nieuwenhuis, J.\ A.\ Tjon \\
\vspace{0.4cm}
{\em Institute for Theoretical Physics, University of Utrecht,\\
Princetonplein 5, 3508 TA Utrecht, the Netherlands.}\\
\vspace {6mm}
Yu.\ A.\ Simonov\\
\vspace{0.4cm}
{\em  Institute for Theoretical and Experimental Physics,\\
117259 Moscow, Russia.}\\
\vspace{0.7cm}
\end{center}
\begin{abstract}
In this work the Feynman-Schwinger representation for the
two-body Greens function is studied. After having given a brief
introduction to the formalism, we report on the first calculations
based on this formalism. In order to demonstrate the validity
of the method, we consider the static case where the mass of one
of the particles becomes very large.
We show that the heavy particle follows a classical trajectory
and we find a good
agreement with the Klein-Gordon result.
\end{abstract}
\section{Theory}
Recently, the Feynman-Schwinger representation (FSR) was presented
\cite{FSR} as a new covariant formalism to calculate the relativistic
two-body Greens function. It was shown that the FSR sums up
all ladder {\em and} crossed diagrams and that it has the correct
static limit if one of the masses of the two particles becomes
very large. Furthermore it was argued that the formulation
was well suited for essentially nonperturbative gauge theories
such as QCD, since the formalism can be set up in a gauge
invariant way and the possibility exists to include a vacuum
condensate in the interaction kernel via the cumulant expansion.
While the formalism neglects all valence particle loops, one has
the possibility to include all self-energy and vertex-correction
graphs as well. In this paper however, we will concentrate on
the static limit.

In order to demonstrate the formalism we consider the $\phi^3$-theory 
for two charged particles $\chi_1$ and $\chi_2$ with masses
$m_1$ and $m_2$ and charges $g_1$ and $g_2$, 
interacting through exchange of a third,
neutral particle $\phi$ with mass $\mu$. The Euclidean action
for this theory is:
\begin{eqnarray}
S_E& =&\int{\rm d}^4x\left[ \chi_i^{\dagger} \left( m_i^2-\partial_{\mu}^2
+g_i\phi\right)\chi_i + \mbox{$\frac{1}{2}$}\phi
\left(\mu^2-\partial_{\mu}^2\right)\phi\right]\\
&\equiv & \int {\rm d}^4x \left[
\chi_i^{\dagger} \Lambda_i \chi_i + \mbox{$\frac{1}{2}$}\phi
\left( \mu^2 - \partial_{\mu}^2 \right) \phi\right]\nonumber
\end{eqnarray}
where summation over the index $i$ is implied. The Greens
function is defined as the transition probability from the
initial state $\Psi_{\rm i}=\chi^{\dagger}_1(x_1) \chi_2(x_2)$ to the
final state $\Psi_{\rm f}=\chi_1(y_1)\chi_2^{\dagger}(y_2)$:
\begin{displaymath}
\hspace{-6cm}
G(y_1,y_2|x_1,x_2) \propto  \int {\cal D}\chi_1{\cal D}\chi_2
{\cal D}\phi \;\;\Psi_{\rm f}\Psi_{\rm i} \;\;{\rm e}^{-S_E}
\end{displaymath}
\begin{equation}
\propto  \int {\cal D}\phi \; \left( \det \Lambda_1(y_1,x_1)
\Lambda_2(y_2,x_2)
\right)^{-\frac{1}{2}}\;\Lambda_1^{-1}(y_1,x_1)\Lambda_2^{-1}(y_2,
x_2)\;\;{\rm e}^{-\frac{1}{2}\int {\rm d}^4x \;\phi \left( \mu^2 -
\partial_{\mu}^2 \right) \phi}\label{2}
\end{equation}
Next, in order to get the sum of all generalized ladder
graphs, we
neglect the determinant in (\ref{2}). This corresponds to neglecting all the
$\chi_1$- and $\chi_2$-loops and is often called the `quenched
approximation'. We now wish to rewrite (\ref{2}) in such a way that we
can perform the integration over the remaining field $\phi$ as well.
To this end we exploit the Feynman-Schwinger representation for
$\Lambda^{-1}$:
\begin{eqnarray}
& & \Lambda^{-1}(y,x) \;\;=\;\; \int_0^{\infty} {\rm d}s \left\langle y \left| 
{\rm e}^{-s(m^2-\partial_{\mu}^2 + g\phi)}\right| x \right\rangle 
\nonumber \\
& = & \left. \int_0^{\infty}{\rm d}s 
\lim_{N\rightarrow \infty} 
\left( \frac{N}{2\pi s}\right)^{2N} \int
\prod_{n=1}^{N-1} {\rm d}^4 z_n \;
{\rm e}^{-m^2 s-\frac{N}{4s}\sum_{i=1}^{N} \left( z_i - z_{i-1} 
\right)^2 - \frac{gs}{N}\sum_{i=1}^{N}\phi \left(\frac{1}{2}(z_i+
z_{i-1})\right)}\right|_{z_0=x}^{z_N=y}\nonumber\\
&  =&  \int_{0}^{\infty} {\rm d}s \int
\left( {\cal D}z\right)_{xy} {\rm e}^{-m^2s-\frac{1}{4s}\int_{0}^{1}
\dot{z}_{\mu}^2(\tau ) {\rm d}\tau - gs \int_{0}^{1}\phi(z(\tau )) {\rm
d}\tau}
\label{FS}
\end{eqnarray}
Integrating over the $\phi$-field is now straightforward and
yields:
\begin{displaymath}
\hspace{-10cm}G(y_1,y_2|x_1,x_2)\;\;  \propto 
\end{displaymath}
\begin{equation}
\int_{0}^{\infty}{\rm d}s_1
\int_{0}^{\infty}{\rm d}s_2 \int \left( {\cal D}
z_1\right)_{x_1y_1} \left( {\cal D}z_2\right)_{x_2y_2}
{\rm e}^{-m_1^2s_1-m_2^2s_2-\frac{1}{4s_1}\int_0^1 \dot{z}_{1,\mu}^2
(\tau_1 )
{\rm d}\tau_1 - \frac{1}{4s_2}\int_0^1 \dot{z}_{2,\mu}^2 
(\tau_2 ){\rm d}
\tau_2} \left\langle W_{\phi} \right\rangle
\label{green}
\end{equation}
In case we leave out the self-energy contribution and the vertex-corrections, 
the Wilson loop $\left\langle W_{\phi}
\right\rangle$ is simply given by:
\begin{eqnarray}
\left\langle W_{\phi} \right\rangle & = & \exp \left[ -\frac{g_1g_2 s_1
s_2 }{N^2}\sum_{j,k=1}^{N} \Delta \left( \mbox{$\frac{1}{2}$}
\left( z_{1,j}+z_{1,j-1}-z_{2,k}-z_{2,k-1}\right)\right)\right]
\nonumber \\
& = & \exp \left[ -g_1g_2s_1s_2 \int_0^1\int_0^1 
\Delta \left(z_1 (\tau_1 ) - z_2 (\tau_2 ) \right) {\rm d}\tau_1
{\rm d}\tau_2\right]
\label{wilson}
\end{eqnarray}
where $\Delta$ is the free scalar two-point function:
\begin{equation}
\Delta (z) =
\frac{\mu}{4\pi^2 \left| z \right|}
K_1 ( \mu \left| z \right| )
\end{equation}
This formulation for the Greens function has the great
advantage that it is essentially 
a quantummechanical 
one in which the two valence particles interact via a nonlocal 
interaction.
One can show that (\ref{green}) and (\ref{wilson}) sum up all
ladder and crossed diagrams. Furthermore, if $m_2 \rightarrow 
\infty$ then $G$ obeys the following Klein-Gordon equation:
\begin{equation}
\left( \partial_{\mu}^2 - m_1^2 - \frac{g_1g_2}{2m_2}V(\left|
{\bf r}\right|)\right)G(r)=\delta^{(4)} (r)
\label{KG}
\end{equation}
with
\begin{displaymath}
V( \left| {\bf r} \right| ) = \int_{-\infty}^{\infty} {\rm d}t
\;\; 
\Delta \left( \sqrt{ \left| {\bf r}\right|^2 + t^2} \right)
\end{displaymath}
which reduces to the instantaneous Coulomb interaction for 
the case $\mu \rightarrow 0$.
\section{Results}
For a Euclidean theory it is known that for large times 
$T=\frac{m_1(y_{1,4}-x_{1,4})+m_2(y_{2,4}-x_{2,4})}{m_1+m_2}$
the contributions to $G$ are dominated by the lowest lying
states and that they fall off exponentially,
$G=\sum_i c_i \exp (-E_i T)$. Hence, we can obtain the ground
state energy by observing that:
\begin{equation}
E_0 = -\lim_{T\rightarrow \infty} \frac{{\rm d}}{{\rm d}T}\log G =
-\lim_{T\rightarrow \infty} \left(\frac{{\rm d}}{{\rm d}T}G\right)/G
\end{equation}
The time derivative with respect to $T$ can be done explicitely. 

We project the Greens function on a complete set of total
and angular momentum states $\left| {\bf P}, l, m \right\rangle$, 
but due to the nonlocal interaction (\ref{wilson}) this does {\em not}
imply that the degrees of freedom that are associated with the 
generators of the symmetries (${\bf P}$, $L^2$, $L_z$), can be
integrated out. 

For all our calculations we used the Metropolis Monte Carlo algorithm
to perform the integrations over $s_1$, $s_2$ and all the 
coordinates.
The value of $N$ in (\ref{FS}) and (\ref{wilson}) that we needed to get reasonably stable results
was typically 25, so that effectively we had to perform a 200-dimensional
integral. Convergence was usually reached after $3\cdot 10^8$ points
which took approximately 30 hours of CPU-time on our fastest workstation.
\begin{figure}
\epsffile{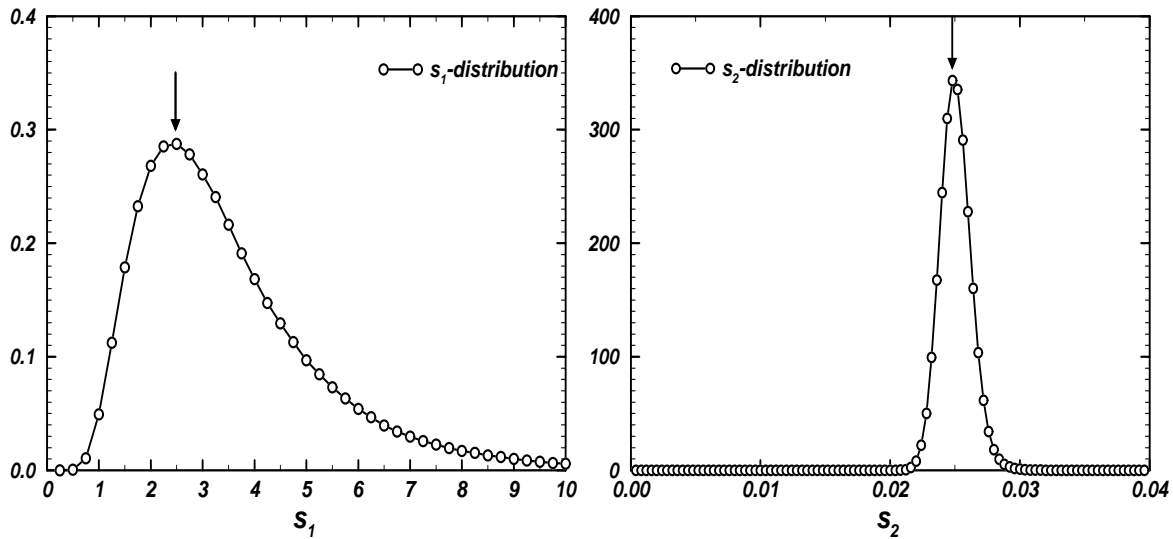}
\caption{{\sl Normalized distributions of $s_1$ and $s_2$. The
arrows indicate the classical values.}}
\label{fig-s}
\end{figure}
In order to demonstrate that it is indeed possible to determine the actual
groundstate within this formalism, we have investigated the case
of one light and one heavy particle interacting through exchange
of a massless third particle ($m_1=1$, $m_2=100$ and $\mu=0$). For
the strength of the coupling we took $g^2\equiv -g_1g_2=4500$.
The distributions of the coordinates of particle 2 are confined to a
very narrow region around their classical values, demonstrating that
the heavy particle indeed follows a classical trajectory. Particle 1 shows
much more quantum behavior; its distributions have finite
widths.
The normalized distributions of $s_1$ and $s_2$ are shown in figure
\ref{fig-s}. Note the difference in scales between both distributions.

\begin{figure}
\epsfxsize=16cm
\epsfysize=12cm
\epsffile{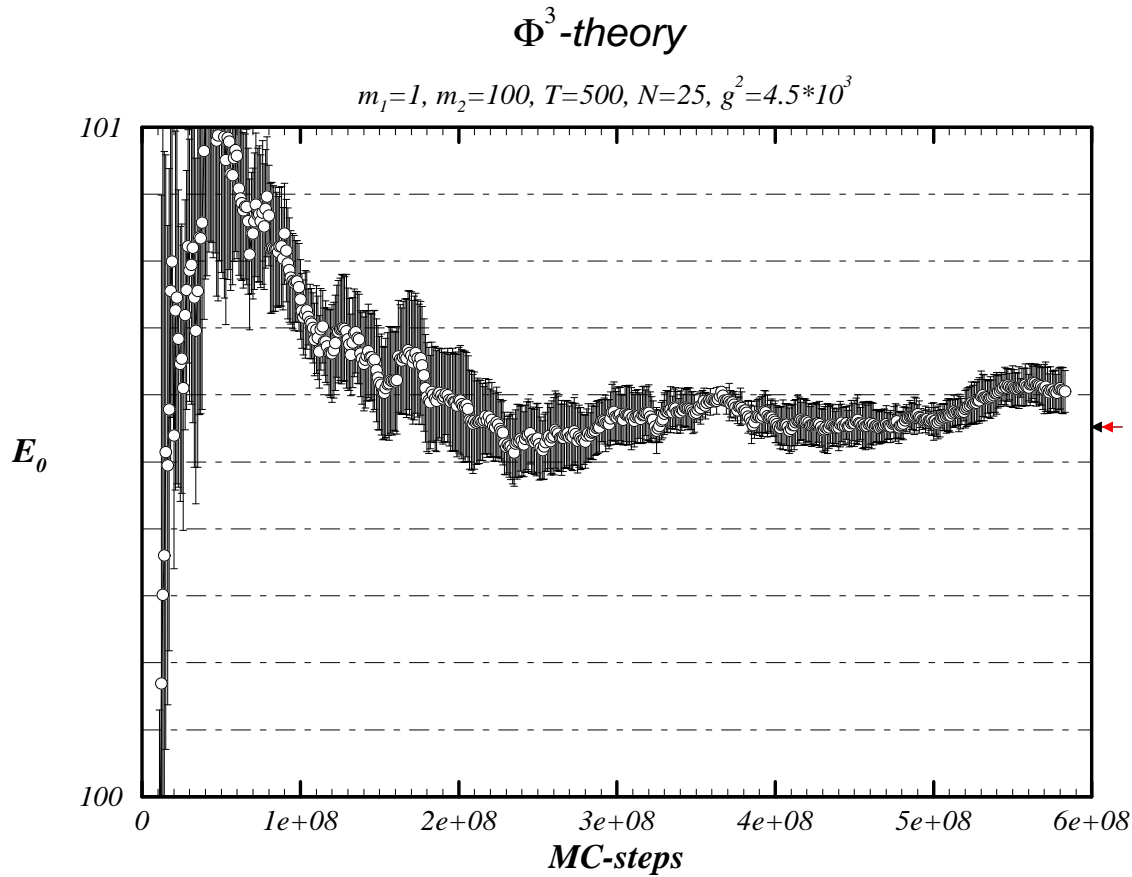}
\caption{{\sl Bound-state mass in the static limit
as a function of the number of Monte Carlo
points. $T$ is given in units of the inverse total mass.}}
\label{fig1}
\end{figure}

In figure \ref{fig1} we show the average of 5 runs as a function of 
the number of Monte Carlo points.
The exact result for $m_2 \rightarrow \infty$,
obtained by solving the Klein-Gordon equation (\ref{KG}), is $E_0=
100.55$ and is indicated by the arrow on the right hand side of
the figure. The FSR-result fluctuates around this value
and the average over the last $4\cdot 10^8$ Monte Carlo
points is $E_0 = 100.57 \pm 0.02$. This clearly 
demonstrates the feasibility of the method. 

We have also studied in detail
the situation where $m_1=m_2=1$ and $\mu =0$. This case
is well suited to test the ladder approximation since the
solution of the
Bethe-Salpeter equation in the ladder approximation is
known exactly
\cite{WC,original1,original2}. For values of the coupling constant such that there
is a binding energy of roughly 10\% of the mass of the
constituents, we find that the full results lie substantially
below the ladder predictions.

In conclusion, we have demonstrated that the FSR is well suited
for going beyond the ladder theory in the study 
of composite systems
in field theory. In particular, ground state properties such
as the binding energies and bound-state wave functions can
readily be determined with this method.

\end{document}